# Identifying and Modeling Security Threats for IoMT Edge Network using Markov Chain and Common Vulnerability Scoring System (CVSS)

Maha Allouzi
mallouzi@kent.edu
Computer Science department
Kent State University

Javed I. Khan
javed@kent.edu
Computer Science department
Kent State University

**Abstract:**

In this work, we defined an attack vector for networks utilizing the Internet of Medical Things (IoMT) devices and compute the probability distribution of IoMT security threats based on Markov chain and Common Vulnerability Scoring System (CVSS). IoMT is an emerging technology that improves patients' quality of life by permitting personalized e-health services without restrictions on time and site. The IoMT consists of embedded objects, sensors, and actuators that transmit and receive medical data. These Medical devices are vulnerable to different types of security threats, and thus, they pose a significant risk to patient's privacy and safety. Because security is a critical factor for successfully merging IoMT into pervasive healthcare systems, there is an urgent need for new security mechanisms to prevent threats on the IoMT edge network. Toward this direction, the first step is defining an attack vector that an attacker or unauthorized user can take advantage of to penetrate and tamper with medical data. In this article, we specify a threat model for the IoMT edge network. We identify any vulnerabilities or weaknesses within the IoMT network that allow unauthorized privileges and threats that can utilize these weaknesses to compromise the IoMT edge network. Finally, we compute the probability distribution of IoMT threats based on the Markov transition probability matrix.

## I. INTRODUCTION:

The Internet of medical things (IoMT) is an evolving technology intending to improve patients' quality of life by enabling personalized e-health services without time and location constraints. However, IoMT devices (e.g., medical sensors and actuators) that compose the fundamental elements of the IoMT edge network and form what is referred to as Wireless Body Area Network (WBAN) are susceptible to various types of security threats, and thus, they might cause a significant risk to patient's privacy and safety. Based on that and the fact that security and privacy are critical factors for the successful adaptation of IoMT technology into pervasive healthcare systems, there is a severe need for new security mechanisms to preserve the security and privacy of the IoMT edge network (WBAN) as utilizing existing authentication solutions to the Internet of Medical Things (IoMT) is not straightforward because of highly dynamic and possibly unprotected environments, and untrusted supply chain for the IoT devices.

So to better understand the specification of the suitable IoMT authentication system, first we need a complete understanding of existing and potential threats to the IoMT edge network environment. Thus, in this article, we first provide a categorization of security threats to the edge network environment based on the significant security objectives that they target. Subsequently, we present a categorization of security countermeasures against threats to IoMT edge networks; this provided a foundation for developing a proper and secure authentication and authorization access control system to protect IoMT edge networks against internal and external threats.

This paper is organized as the following: section II illustrates the IoMT architecture and components to be considered in the threat modeling; section III identifies the weaknesses and vulnerabilities in the IoMT environment; section IV identifies IoMT edge network security threats; section V tabulate the correlation between IoMT security threats and vulnerabilities; section VII provides the distribution of security threat probability using Markov chain and Common Vulnerability Scoring System (CVSS), and finally section VII provides a conclusion and future works.

## II. IoMT SYSTEM ARCHITECTURE

IoMT enables IoT-based healthcare systems to monitor various vital signs such as insulin level, ECG, heart rate, and blood pressure.

The main component of the IoMT-based healthcare system is the IoMT edge network, as shown in Fig.1, it integrates many constrained devices capable of sensing the neighboring environment, obtaining data (such as heart rate, insulin level, etc.), and delivering them to a sink node, also called network coordinator, through a low power and short-range wireless communication technology. The sink node is attached to the gateway that acts as a resource server responsible for receiving and forwarding the medical data based on network availability to either (i) a cellular-based station, through a long-range wireless technology (e.g., 4G/5G) or (ii) a router, through short coverage communication protocols, such as Bluetooth and 6LoWPAN, or Wi-Fi so that medical data will reach, over the Internet, the cloud platform service at the healthcare provider side for future data processing and storage [8,9].

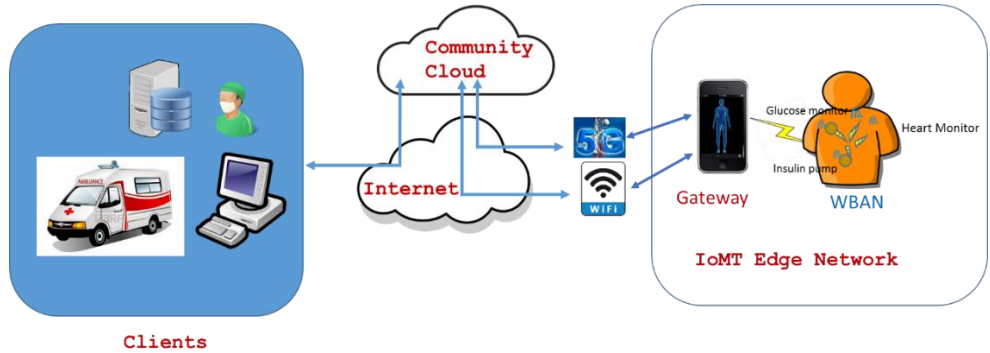

Fig. 1 IoMT System Architecture

### A. RESEARCH PROBLEM

The main goal of this research work is to identify an attack vector for the IoMT edge network. To accomplish our primary research goal, we split the research problem into the following particular objectives.

- First, we identify device-level vulnerabilities and indicate what security vulnerabilities are created at the network level due to the IoMT devices' deployment in the system.
- Second, we identify and classify security threats targeting the IoMT edge network based on the security requirements they intend to compromise.
- Each vulnerability creates one or more security threats that introduce a penetration point(s) in an IoMT network; therefore, our third research problem is to develop correlations between different threats and exploited vulnerabilities.
- Finally, we used Markov Chain and Common Vulnerability Scoring System (CVSS) to distribute IoMT security threat probabilities.

## III. IoMT EDGE NETWORK VULNERABILITIES

This section considers some common vulnerabilities IoMT devices have, which can be exploited by different attacks. Vulnerabilities are typically defined as weaknesses that allow an attacker to perform unauthorized action, e.g., accessing medical data without proper privileges. In this section, we categorized IoMT edge network vulnerabilities based on the source of the vulnerability as; network-based vulnerabilities and device-based vulnerabilities.

### A. IOMT DEVICE VULNERABILITY:

(1) **Weak passwords ($V_1$):** it is a common issue with IoMT devices. When setting passwords for their devices, users often choose an easy password or even decline using a password for their device entirely; this makes it easy for attackers to either guess or brute force their way into a user's account or device [3].

(2) **Command injection flaw ($V_2$)**: injection flaws allow attackers to send malicious code through an application to another system. These attacks include calls to the operating system via system calls, the use of external programs via shell commands, and calls to backend databases via SQL (i.e., SQL injection) [3].

(3) **Insecure web interface ($V_3$)**: An insecure web interface can be present when issues such as lack of account lockout or weak credentials are present. An example of this vulnerability could not use proper sandboxing of a user session; this can allow individuals to utilize JavaScript injection in a cross-site-scripting (XSS) attack. Similarly, input fields may not be properly sanitized, which can allow the further injection of commands. Insecure interfaces can deploy intrusive methods of tracking users and leak data. A phishing attack and steal account logins and credentials [3].

(4) **No account lockout ($V_4$):** When attempting to log in to a website, app, or device, many services have no account lockout feature when incorrect passwords are entered; this allows attackers to continuously try any variation of passwords, pins, or patterns until they can figure out the correct one. They can either do this manually or set up software that performs brute force or a dictionary attack [3].

(5) **Unlimited resource allocation ($V_5$)**: Erroneous modeling of resource usage can lead to overbooking or overprovisioning [3].

B. NETWORK VULNERABILITIES:

   a. **Unencrypted services ($V_6$):** data confidentiality is a vital component of information security. Strong encryption of communications and data storage can prevent unauthorized parties from gaining intelligence in the event of an attack. If an organization chooses to use weaker encryption standards for their services, they are vulnerable to information leakage, eavesdropping, and other types of surveillance. This is because the devices send packets over the network to transmit data which can be easily intercepted and read via packet sniffers [3].

   b. **Open ports ($V_7$)**: When data is transmitted across a network, it uses specific types of ports depending on its category. By default, all ports are open. However, users can choose to close any unused ports to reduce the attack surface for attackers. Firewalls often provide the option to close most ports that are not commonly used. Still, users can make changes on a case-by-case basis to achieve the best security they can tolerate after configuring their settings [3].

   c. **Insecure network services ($V_8$)**: Legacy network protocols are vulnerable to a variety of attacks. Protocols such as Telnet and Server Message Block (SMB) feature obsolete countermeasures or are not designed for security by lacking default encryption. Devices can be vulnerable to default logins such as login: root with Telnet. SMB also has vulnerabilities that are associated with the ransomware infections, Petya and Notpetya. Network services are made insecure by several aspects, including buffer overflow, open ports, exploitable UDP services, etc. Some examples of attacks that could be used against insecure network services include fuzzing attacks and using open ports to gain access to the network [3].

   d. **Insecure cloud interface ($V_9$):** Insecure connections reduce the expectation of data confidentiality. An insecure cloud interface can use insecure protocols that lack Secure Socket Layer (SSL) encryption features. An insecure application may also allow attackers to eavesdrop upon concurrent users. This can be used to exfiltrate data, cause unintended functionality, or violate user confidentiality. Several different aspects determine whether the cloud interface is secure or not, including lacking account enumeration, not having any account lockout, unencrypted data while traveling over the network, and others. Examples of attacks are an attacker determining if an account is valid by trying to reset the password or using a packet sniffer to find login credentials through unencrypted traffic [3].

   e. **Removal of physical storage ($V_{10}$):** The removal of physical storage is a physical attack vector. The removal of physical storage is specified in a storage pool. If sufficient space does not exist in the storage pool to tolerate this removal, then data loss can be the ending result in which the user is warned about this. If the user configuration allows, then the user should add a replacement physical storage to the pool before removing the old one; this also allows attackers to access user information if the data is left unencrypted [3, 4].

   f. **Missing authorization ($V_{11}$):** Authorization determines whether a user has the necessary privileges to access resources and performs actions. This is generally referred to as Role-Based Access Control (RBAC). The RBAC mechanism provides a way of restricting system access and activities to authorized users. In RBAC, access control is defined around roles and privileges. In the absence of an appropriate authorization mechanism, once a user is authenticated, they can perform actions however damaging they are [12].

## IV. IoMT EDGE NETWORK SECURITY THREATS

This section provides a classification of the security threats aiming at the IoMT edge network based on the security objectives that they intend to compromise. First, we list the security objectives in the IoMT edge network, then we present types of attacks that can be potential attacks against IoMT edge networks [11].

### A. SECURITY REQUIREMENTS FOR IOMT EDGE NETWORK:

This section identifies the six primary security objectives concerning the IoMT environment.

   a. CONFIDENTIALITY

Assures that confidential medical data is not exposed or disclosed to unauthorized entities. In the context of the IoMT network, confidentiality refers to the protection of a patient's medical information, shared with a therapist, a physician, or medical staff, from being disclosed to unauthorized third parties that can harm the patient or use this medical information in an inappropriate manner [6].

   b. INTEGRITY

Assures that data has not been modified or altered in an unauthorized manner. Concerning the IoMT edge network, integrity preserves the accuracy of patient's records [5].

   c. AUTHENTICATION

Authentication applies to both entities and message authentication. Entity authentication or identification is the process by which one communicating entity is assured of the claimed identity of another entity involved in the interaction and that the latter has participated. On the other hand, message authentication is the process by which an entity is verified as the original source of given data generated at some time in the past. Nowadays, there is a trend toward lightweight authentication protocols. Many IoT devices are resource constraints and do not have enough memory and CPU power to execute the cryptographic operations required for traditional authentication protocols.

   d. AUTHORIZATION

Authorization guarantees that only permitted entities can obtain access to specific network services or resources, such as a medical IoT device or collected medical data of a patient. For instance, only trusted expertise parties are granted permission to perform a given action, such as issuing commands to medical IoT devices or updating the medical IoT device software. Access control is a standard security technique that ensures authorization.

   e. AVAILABILITY

Ensures that systems are available to provide services and work properly and services are not denied to authorized users. Therefore, medical data are always accessible and useable upon demand by a legitimate entity. In the context of the IoMT edge network, it is of significant importance to ensure the availability of device and network resources when a patient needs care services without disruptions.

B.  TYPICAL THREAT TYPES IN IOMT EDGE NETWORK:

The growing number of resource-constrained medical devices connected to the Internet over wireless networks leads to security breaches by malicious actors who exploit possible system vulnerabilities to launch attacks and gain access to confidential information or affect extracted results and device operations. This section presents a brief description of generalized types of attacks that can be potential attacks against IoMT edge networks.

a. **Man-in-the-middles attacks ($A_1$)**: This kind of active attack occurs when a malicious actor interferes in the communication between two authenticated entities (e.g., the claimant and verifier of the authentication protocol), intercepting, compromising, or even concealing messages exchanged to each other. The attacker may selectively alter the communicated data to masquerade as one or more of the legitimate entities involved [10].

b. **Spoofing attacks ($A_2$):** Spoofing is another possible security threat for IoMT networks. It allows the attacker to appear to be a legitimate user as opposed to showing their actual device. The attack is executed by sending messages to the IoMT resource server, in which the sender attempts to masquerade as a trusted source. This is done to gain access to the target's sensitive information, such as medical data.

c. **Traffic analysis attacks ($A_3$)**:  A type of passive attack in which an attacker gains knowledge of the transmitted information by inference from observable data flow characteristics. The information may not be directly available, for instance, when the data are encrypted. These characteristics may include the identities and locations of the involved entities (i.e., sources and destinations) of the data flow and the flow's presence, absence, amount, direction, frequency, and duration.

d. **Masquerading attacks ($A_4$):** A type of active attack whereby unauthorized entities illegitimately pose as authorized entities to gain a greater privilege to the system than what they are authorized for. Moreover, the attacker may perform a malicious action by illegitimately posing as an authorized entity. For instance, the attacker may steal the user's terminal device (e.g., user's smartphone) login credentials and gain unauthorized privileges to access stored confidential health data by masquerading the legitimate user. An attacker could pretend to be a legitimate user to gain access to services that the IoMT device provides via the insertion of rogue devices, or an attacker is apparently presented as an IoMT device to offer fake services to users. The second case is hazardous in the healthcare sector, where the services provided by IoMT devices are life-dependent for many patients.

e. **Physical attacks ($A_5$)**:  Physical attacks are concentrated on the physical layer and the devices themselves. For instance, an adversary may change the behavior or structure of devices involved in the IoMT edge network by leading the system to hardware failure. Examples of physical attacks include device capture, tampering, invasive hardware attacks, side-channel attacks, and reverse engineering attacks [5].

f. **Malware attacks ($A_6$)**:   An attacker designs and operates malicious software or firmware to violate a system's security. This software or firmware is often covertly inserted into another program. It intends to destroy data, run destructive or intrusive programs, or otherwise compromise the privacy, accuracy, or reliability of the system's data, applications, or the operating system as a whole. Typical means for malware attacks include worms, virus programs, malicious mobile code, trojan horses, rootkits, or other code-based malicious entity that successfully infects a system

g. **Eavesdropping attacks ($A_7$):** Attacks that take advantage of insecure network communications to interfere between the communicating entities to access data as it is being sent or received by its users, such as smartphones or sensor nodes, without their consent.

h. **Message fabrication/modification/replay attacks ($A_8$)**: Finally, in message fabrication/modification and replay attacks, the adversary can construct, change, or resend, respectively, already transmitted messages between legitimate entities with the intent of producing an unauthorized effect or gaining unauthorized access

i. **Social Engineering attacks ($A_9$)**: Social Engineering is the process of manipulating people and their devices to obtain sensitive information such as social security numbers and bank account information etc. Social engineering involves social interactions with the intent of exploiting an individual's trust; this relates to IoT devices because if an attacker can intercept information from them, they can use it to connect or manipulate their targets. This can be something as simple as finding what bands they're interested in and pretending they share the same interests, or something more involved by gaining access to their private information and using that to coerce them into acting however you want them to.

j. **Ransomware attacks ($A_{10}$):** Ransomware is a method where an attacker locks a user out of their device and demands payment to unlock it. They can either threaten to simply delete the user's files if they don't comply or threaten to upload them to the Internet if it is private information or media that the user wouldn't want everybody to have access to online. This attack targets not only individuals and their devices but also larger companies and corporations as well [7]. Recently, larger entities such as trains and banks have become victims of ransomware attacks to extort them into paying large sums of money. Ransomware infections have become more common with recent threats such as Petya and WannaCry.

k. **Elevation of privilege ($A_{11}$):** Elevation of Privilege (EoP) is the ability of an attacker to gain essential, limited access to a system, and increase their privileges, what a user is permitted to do within a system afterward [5]. This would allow them to access information and systems that they cannot otherwise access by promoting themselves to higher-level accounts that can make more changes to the system. For example, an attacker with a privileged set of "read-only" permissions elevates the set to include "read and write." Therefore, an EOP attack is when a user receives privileges they are not entitled to.

l. **Denial-of-service attacks ($A_{12}$):** It is one of the most common types of attacks. It occurs whenever an attacker floods a server with different types of packets, causing it to either crash from the load or rejecting authentic users from the service. The attacks can be further broken down into three main types. These include volume-based, protocol, and application layer. Volume-based attacks work by saturating the bandwidth until it can't cope with the high traffic levels, such as through NTP Amplifications. Protocol attacks work by consuming server resources via fragmented packets. The server waits for the following pieces that are supposed to come paired with the original messages, such as ICMP and SYN floods. Application layer attacks intend to crash the web server such as HTTP and SIP flood [10]. Whenever a botnet commits an attack, it is known as a Distributed Denial of Service, or DDoS attack.

## V. THE CORRELATION BETWEEN IOMT SECURITY THREATS AND VULNERABILITIES

A security attack usually exploits one or more vulnerabilities in the components of a system to compromise it. The correlation between security vulnerability and these recognized threats is thus essential for the threat model. In this section, we tabulate the relationship between the IoMT threats and vulnerabilities. And this correlation will be used to quantify the probability for each threat.

Table I the correlation between IoMT security threats and exploited vulnerabilities

| Security Requirement violated | STRIDE Threats | Attacks | Vulnerability Exploited |
|---|---|---|---|
| **Authentication** | Spoofing | **A1:** Man-in-the- middles attacks | $V_3, V_6, V_8, V_9, V_{10}$ |
| | | **A2:** Spoofing attacks | $V_3, V_6, V_8, V_9, V_{10}$ |
| | | **A4:** Masquerading attacks | $V_3, V_6, V_8, V_9, V_{10}$ |
| **Authorization** | Elevation of Privileges | **A12**: Elevation of privilege | $V_7$ |
| | | **A6:** Malware attacks | $V_2, V_8, V_9$ |
| **Integrity** | Tampering | **A3:** Traffic analysis attacks | $V_6, V_8, V_9$ |
| | | **A5:** Physical attacks | $V_2, V_{10}$ |
| | | **A8:** Message fabrication/modification/replay attacks | $V_6$ |
| **Availability** | DoS threats | **A10:** Ransomware | $V_1, V_2, V_6, V_8, V_9$ |
| | | **A12:** Denial-of-service attacks | $V_4, V_9, V_{11}$ |
| **Information Disclosure** | Confidentiality | **A7:** Eavesdropping attacks | $V_6, V_3, V_8, V_9$ |
| | | **A9:** Social engineering | $V_{10}$ |
| | | **A6:** Malware attacks | $V_2, V_9$ |

## VI.  MARKOV MODEL FOR SUCCESSFUL ATTACKS

A security threat is a stochastic process that can be modeled as a Markov chain. The probability of transition from a state to another is based on the vulnerabilities present in the current state. An attacker exploits various vulnerabilities to reach a security Threat state (T) and eventually reaches the Attacked state (A). We model the process using a quantifiable state, namely the Safe state (S), the Threat state (T), and the Attack state (A). Fig.2 depicts the proposed Markov chain for modeling security threats and attacks with state transition probabilities. α denotes the transit probability from state S to state T, ß denotes the transit probability from state T to state S, μ denotes the transit probability from state T to state A. ε denotes the transit probability from state A to state S. this model takes into the consideration all factors of an attack, including countermeasures.

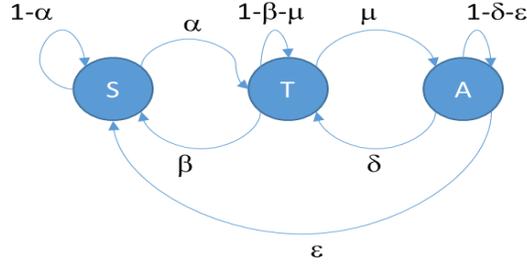

Fig. 2 Markov model with countermeasures

However, in this work, we don't take into consideration the application of countermeasures and studied only the probability of possible attacks, as our primary goal is to examine the impact of security threats on attacked system and how attacker exploits system vulnerabilities to change the state of the system from Safe state to Attack state. This model is shown in Fig. 3 and is a simplified model of Fig. 2.

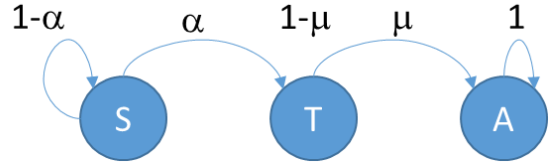

Fig. 3 Markov model without considering the countermeasures

Our goal is to find the transition probability from the Safe state to the Threat state. Using the Chapman-Kolmogorov equation, we can find the transient probabilities between two states after several jumps-steps. Since the distribution of probability on the state of a Markov chain is discrete, and this space is homogeneous, it can be showed by matrix multiplication. Chapman-Kolmogorov equation can be used as follows:

$$P_{ij}^{(n+m)} = \sum_{k=0}^{\infty} p_{ik}^{(n)} P_{kj}^{(m)} \qquad (1)$$

Where P is the probability matrix of transition state space. $P_{ij}^{(n+m)}$ The transition probability from state i to state j after (m + n) steps with an intermediate stop in state k after n steps; then sum over all possible k values.

## VII. DISTRIBUTED OF SECURITY THREAT PROBABILITY

To compute the distribution of security threats probabilities based on the Markov chain and transit probability matrix, we first modeled the IoMT security threats as a transition graph shown in Fig.4. This model has 14 states, including Safe state (S), Attack state (A), and 12 Threats states ($T_1...T_{12}$). A safe state is a state where the system has no threats. The Attack state is a state when the system fails to meet one or more of the security requirements listed in section (IV-A) due to exploiting the security vulnerability of a specific threat.

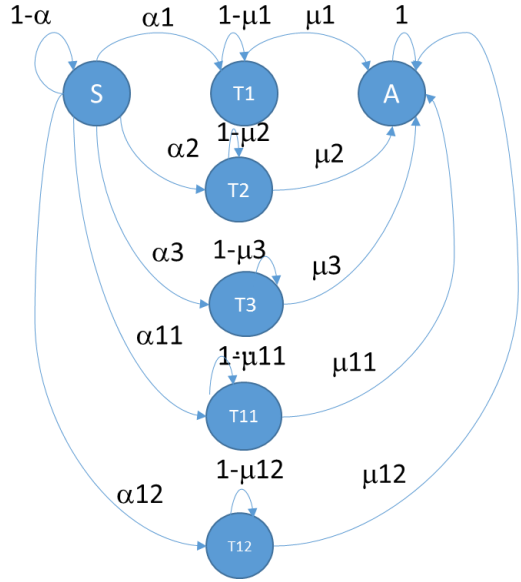

Fig. 4 IoMT threat Model

From this model we build a transition probability matrix. The probability of each attack path is considered as the probability of moving from state Safe to Sate Attack. And this happen when attackers leverage the system vulnerabilities and launch a successful attack on the IoMT edge network. The threat model in Fig.4 can be represented as a transition probability matrix which is shown in fig.5.

$$P = \begin{bmatrix} 1-\alpha & \alpha 1 & \alpha 2 & \alpha 3 & \alpha 4 & \alpha 5 & \alpha 6 & \alpha 7 & \alpha 8 & \alpha 9 & \alpha 10 & \alpha 11 & \alpha 12 & 0 \\ 0 & 1-\mu 1 & 0 & 0 & 0 & 0 & 0 & 0 & 0 & 0 & 0 & 0 & 0 & \mu 1 \\ 0 & 0 & 1-\mu 2 & 0 & 0 & 0 & 0 & 0 & 0 & 0 & 0 & 0 & 0 & \mu 2 \\ 0 & 0 & 0 & 1-\mu 3 & 0 & 0 & 0 & 0 & 0 & 0 & 0 & 0 & 0 & \mu 3 \\ 0 & 0 & 0 & 0 & 1-\mu 4 & 0 & 0 & 0 & 0 & 0 & 0 & 0 & 0 & \mu 4 \\ 0 & 0 & 0 & 0 & 0 & 1-\mu 5 & 0 & 0 & 0 & 0 & 0 & 0 & 0 & \mu 5 \\ 0 & 0 & 0 & 0 & 0 & 0 & 1-\mu 6 & 0 & 0 & 0 & 0 & 0 & 0 & \mu 6 \\ 0 & 0 & 0 & 0 & 0 & 0 & 0 & 1-\mu 7 & 0 & 0 & 0 & 0 & 0 & \mu 7 \\ 0 & 0 & 0 & 0 & 0 & 0 & 0 & 0 & 1-\mu 8 & 0 & 0 & 0 & 0 & \mu 8 \\ 0 & 0 & 0 & 0 & 0 & 0 & 0 & 0 & 0 & 1-\mu 9 & 0 & 0 & 0 & \mu 9 \\ 0 & 0 & 0 & 0 & 0 & 0 & 0 & 0 & 0 & 0 & 1-\mu 10 & 0 & 0 & \mu 10 \\ 0 & 0 & 0 & 0 & 0 & 0 & 0 & 0 & 0 & 0 & 0 & 1-\mu 11 & 0 & \mu 11 \\ 0 & 0 & 0 & 0 & 0 & 0 & 0 & 0 & 0 & 0 & 0 & 0 & 1-\mu 12 & \mu 12 \\ 0 & 0 & 0 & 0 & 0 & 0 & 0 & 0 & 0 & 0 & 0 & 0 & 0 & 1 \end{bmatrix}$$

Fig. 5 transition probability matrix

$\alpha$ is denotes the sum of probability of all attack paths from Safe state to Threat State. And $\mu$ is the sum of all probability of all threats states to attack state. When the system is in the Safe state, it will remain in this state with probability (1-$\alpha$) and once the system is in the Attack state, the probability is 1. The probabilities of attack paths from S to T states are $\alpha 1, \alpha 2, \alpha 3 \ldots \alpha 12$. The probabilities of attack paths representing from threats states to the Attack state are $\mu 1, \mu 2, \mu 3 \ldots \mu 12$.

Finally, we compute the transition probability from the Safe state(S) to the Attack state (A) via threats $T_i$. Each attack path represents the path that the attacker will take advantage of to reach the Attack state from the threat state $T_i$ by exploiting the set of vulnerabilities ($V_{ij}$) of each security threat as listed in Table I. for example for the $A_1$: Man in the Middle attack to happen the following vulnerabilities should be exploited {$V_3, V_6, V_8, V_9, V_{10}$}. To quantify this attack distribution, we use the weight of each attack path, an attack path is the path in the transition probability graph from the S state to the A state through one of the 12 threats state $T_i$. CVSS can be used to weight each path from state S to state A. The weight associated with the transition from S to $T_i$ is determined by computing the ratio between vulnerability scores from S to Ti and all vulnerability scores from S to all threats. The transition probabilities (αi) from S to $T_i$ can be calculated as the following:

$$\alpha i = \frac{\sum_j v_{ij}}{\sum_{yz} v_{yz}} * \alpha \quad (2)$$

Where:

$v_{ij}$: is the vulnerability score for path i.
$v_{yz}$ : is the vulnerability score of all possible attacks.

Similarly, the transition probabilities (μ) from $T_i$ to F can be computed as the following:

$$\mu i = \frac{\sum_j v_{ij}}{\sum_{yz} v_{yz}} * \mu \quad (3)$$

To compute the transient probability S to F via $Ti$,

$$\Pr(X3 = A \mid X1 = S) = \sum_j \Pr(X3 = A|X2 = Tj) \Pr(X2 = Tj|X1 = S)$$

This sum of product is nothing but the definition of the matrix product, and because the Markov chain is homogeneous

$$P^2 (SF) = P^2 = \alpha i * \mu i \quad \ldots\ldots\ldots (4)$$

Table II Vulnerabilities Scores

| Vulnerability | Acronym | Exploitability score/10 |
|---|---|---|
| Weak passwords | V1 | 6.3 |
| Command injection flaw | V2 | 5.8 |
| Insecure web interface | V3 | 9.6 |
| No account lockout | V4 | 5.7 |
| Unlimited allocation of resources | V5 | 2 |
| Unencrypted services | V6 | 6.8 |
| Open ports | V7 | 7 |
| Insecure network services | V8 | 9.6 |
| Insecure cloud interface | V9 | 9.6 |
| Removal of physical storage | V10 | 7.2 |
| Missing authorization | V11 | 8.6 |

*https://nvd.nist.gov/vuln-metrics/cvss/v3-calculator

To calculate the probability distribution of security attacks, we have to determine the Markov transition matrix elements based on the vulnerabilities associated with an attack. From the Safe state S, the total probability that the system moves to one of the threat states is assumed α (α = 0.0318 [2]). We can calculate the transition probability that the system changes from S to T as the ratio of the sum of i vulnerability scores of threats associated with Ti over the total CVSS scores of all threats.

Table II shows the CVSS scores [1] associated with relevant vulnerabilities considered in this paper. According to CVSS, this number is a score out of ten. For example, V8 scores 9.6/10 because the severity of this vulnerability is very high once it is related to IoT data breach vulnerabilities. Besides, to go to state $A_2$ from S, an attacker needs to exploit a certain set of vulnerabilities associated with the security threat state A3, as listed in Table I. In this case, vulnerabilities three, six, eight, nine, and ten will be exploited (see Table 1). Therefore, the number of vulnerability scores for the attack path one is ($W_{A2}$=V3+V6+V8+V9+V10= 43) and the total number of all vulnerability scores from S to any Ti ($WT_1$). We can estimate the transition probability from S to $A_2$ ($α_1 = 43/390 * α = 0.003506$).Fig 6 illustrates the vulnerability score for all the attacks considers in this paper.

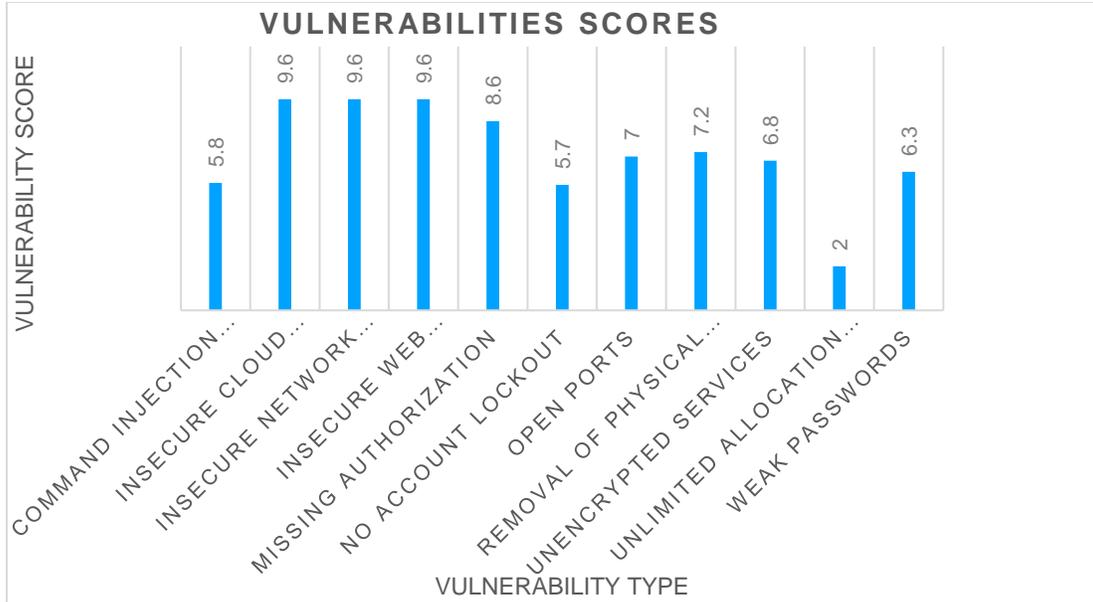

Fig.6 Vulnerabilities score

Similarly, other transition probability from S to $T_i$ will be computed by using (4). We assume that the transition probability from state $T_i$ to A is highly likely with probability $\mu_{iA} = 0.98$ for any attack paths (see Figure 4). By computing $\alpha_i$ and $\mu_{iA}$, the transition probability matrix P is obtained. Using (1) and (4), we have the probabilistic distribution of twelve security threats expressed in Table III and illustrated in Figure 7.

Table III Probability distribution of the 13 security attacks on IoMT Edge Network

|   | Attack | Probability ($*10^{-3}$) |
|---|---|---|
| 1 | $A_1$: Man-in-the- middles attacks | $P_{A1} = \alpha1 * \mu1 = 0.003506*0.98= 3.44$ |
| 2 | $A_2$: Spoofing attacks | $P_{A2} = \alpha2 * \mu2 = 0.003506*0.98=3.44$ |
| 3 | $A_3$: Traffic analysis attacks | $P_{A3} = \alpha3 * \mu3 = .00212*0.98=2.078$ |
| 4 | $A_4$: Masquerading attacks | $P_{A4} = \alpha4 * \mu4 = 0.003506*0.98=3.44$ |
| 5 | $A_5$: Physical attacks | $P_{A5} = \alpha5 * \mu5 = 0.00098*0.98=0.939$ |
| 6 | $A_6$: Malware attacks | $P_{A6} = \alpha6 * \mu6 = 0.00204*0.98=1.99$ |
| 7 | $A_7$: Eavesdropping attacks | $P_{A7} = \alpha7 * \mu7 = 1.132*0.98=1.11$ |
| 8 | $A_8$: Message modification/replay attacks | $P_{A8} = \alpha8 * \mu8 =0.00554*0.98=5.433$ |
| 9 | $A_9$: Social engineering | $P_{A9} = \alpha9 * \mu9 = 0.00587*0.98=5.753$ |
| 10 | $A_{10}$: Ransomware | $P_{A10} = \alpha10 * \mu10 =0.00311*0.98=3.044$ |
| 11 | $A_{11}$: Elevation of privilege | $P_{A11} = \alpha11 * \mu11 =0.003506*0.98= 3.44$ |
| 12 | $A_{12}$: Denial-of-service attacks | $P_{A12} = \alpha12 * \mu12 =0.019*0.98=19.1$ |

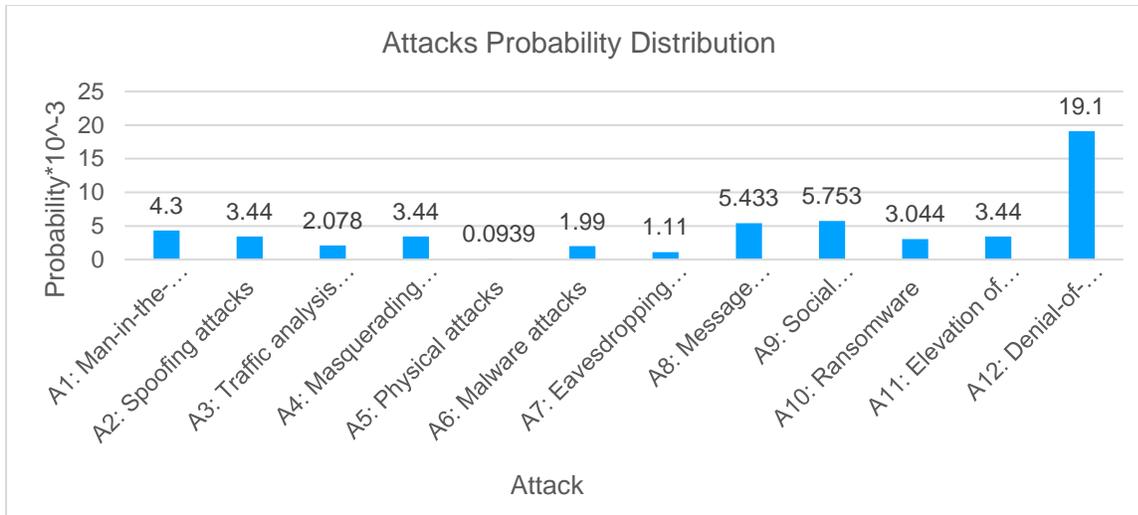

Fig.7 IoMT attacks Probability Distribution

## VIII. CONCLUSION

The attack vector of a network is the sum of all penetration points. In the context of IoMT, insecure IoMT devices can become attack vectors for more significant and life intimidating attacks. Part of minimizing the attack vector is exercising obscurity and limiting the availability of assets to authorized persons. In this work, we find the distribution of security threat probability in the IoMT edge network, which helps to quantify the likelihood of attacks that will exploit these threats. And also give an insight into what vulnerability should have the higher priority of mitigation and security measures. We will use these findings and consider them to secure our Soter system, an authentication, and authorization system for IoMT.